\documentclass{article}

\usepackage{arxiv}

\usepackage[utf8]{inputenc} 
\usepackage[T1]{fontenc}    
\usepackage{hyperref}       
\usepackage{url}            
\usepackage{booktabs}       
\usepackage{amsfonts}       
\usepackage{nicefrac}       
\usepackage{microtype}      
\usepackage{lipsum}
\usepackage{graphicx}
\usepackage{authblk}
\usepackage{float}
\graphicspath{ {./images/} }

\usepackage{amsmath,amssymb}
\usepackage{comment}
\usepackage{mathrsfs}
\usepackage{graphicx,nicefrac}
\usepackage{float}

\usepackage{comment}
\usepackage{scalerel}
\usepackage[overload]{empheq}
\usepackage{textgreek}

\providecommand{\vs}{\mathbf{s}}
\providecommand{\vx}{\mathbf{x}}

\providecommand{\mM}{\mathbf{M}}

\usepackage{algorithm}
\usepackage{algpseudocode}
\usepackage{enumitem}
\usepackage{hyperref}
\usepackage{mathrsfs}
\usepackage{graphicx,nicefrac}

\usepackage[utf8]{inputenc}
\usepackage[english]{babel}

\usepackage[margin=1cm]{caption}
\usepackage{comment}
\usepackage{scalerel}
\usepackage[overload]{empheq}

\usepackage{hyperref}

\usepackage{url,hyperref,lineno,microtype,subcaption}
\usepackage[onehalfspacing]{setspace}

\usepackage{comment}
\usepackage{scalerel}
\usepackage[overload]{empheq}
\usepackage{textgreek}

\providecommand{\vs}{\mathbf{s}}
\providecommand{\vx}{\mathbf{x}}

\providecommand{\mM}{\mathbf{M}}

\begin{document}

\pagestyle{fancy}
\rhead{}

\title{Transient motion classification through turbid volumes via parallelized single-photon detection and deep contrastive embedding}

\author[1]{Shiqi Xu}
\author[2]{Wenhui Liu}
\author[1]{Xi Yang}
\author[3]{Joakim Jonsson}
\author[1]{Ruobing Qian}
\author[4]{Paul McKee}
\author[1]{Kanghyun Kim}
\author[1]{Pavan Chandra Konda}
\author[1]{Kevin C. Zhou}
\author[1]{Lucas Kreiß}
\author[5]{Haoqian Wang}
\author[3]{Edouard Berrocal}
\author[4]{Scott Huettel}
\author[1,*]{Roarke Horstmeyer}

\affil[1]{Department of Biomedical Engineering, Duke University, Durham, NC, USA, 27708}
\affil[2]{Department of Automation, Tsinghua University, Beijing, China, 100084}
\affil[3]{Division of Combustion Physics, Department of Physics, Lund University, Sweden,22100}

\affil[5]{Tsinghua Shenzhen International Graduate School, Tsinghua University, Shenzhen, China, 518055}

\affil[*]{roarke.w.horstmeyer@duke.edu}

\maketitle

\begin{abstract}
Fast noninvasive probing of spatially varying decorrelating events, such as cerebral blood flow beneath the human skull, is an essential task in various scientific and clinical settings. One of the primary optical techniques used is diffuse correlation spectroscopy (DCS), whose classical implementation uses a single or few single-photon detectors, resulting in poor spatial localization accuracy and relatively low temporal resolution. Here, we propose a technique termed \emph{\textbf{C}lassifying \textbf{R}apid decorrelation \textbf{E}vents via \textbf{P}arallelized single photon d\textbf{E}tection (CREPE)}, a new form of DCS that can probe and classify different decorrelating movements hidden underneath turbid volume with high sensitivity using parallelized speckle detection from a $32\times32$ pixel SPAD array. We evaluate our setup by classifying different spatiotemporal-decorrelating patterns hidden beneath a 5mm tissue-like phantom made with rapidly decorrelating dynamic scattering media. Twelve multi-mode fibers are used to collect scattered light from different positions on the surface of the tissue phantom. To validate our setup, we generate perturbed decorrelation patterns by both a digital micromirror device (DMD) modulated at multi-kilo-hertz rates, as well as a vessel phantom containing flowing fluid. Along with a deep contrastive learning algorithm that outperforms classic unsupervised learning methods, we demonstrate our approach can accurately detect and classify different transient decorrelation events (happening in 0.1-0.4s) underneath turbid scattering media, without any data labeling. This has the potential to be applied to noninvasively monitor deep tissue motion patterns, for example identifying normal or abnormal cerebral blood flow events, at multi-Hertz rates within a compact and static detection probe.

\end{abstract}

\section{Introduction}
Non-invasive probing and identification of hemodynamic events deep inside tissue, such as cerebral blood flow (CBF), is essential for both clinical and scientific studies. In the past, numerous optical methods have been developed to detect and monitor CBF, such as diffuse optical spectroscopy (DOS)~\cite{gibson2009diffuse}, diffuse optical tomography (DOT)~\cite{durduran2010diffuse}, functional near-infrared spectroscopy (fNIRS)~\cite{ferrari2012brief}, and photoacoustic tomography (PAT)~\cite{wang2016practical}. These methods typically measure the absorption change caused by blood oxygenation, which is correlated with blood flow change. Recent extension of these methods can probe even deeper into tissue by time-gating multi-scattered light from non-superficial layers ~\cite{torricelli2014time}, which can also be implemented in the frequency domain using polychromatic measurements~\cite{kholiqov2020time}. 

Instead of looking at the absorption change, another class of techniques attempt to measure the dynamics directly by recording the temporal fluctuations of scattered light, among which established techniques are optical coherence tomography angiography (OCTA)~\cite{spaide2018optical} and laser speckle contrast imaging (LASCI)~\cite{briers2013laser}. While there are impressive demonstrations using these methods to create microscopic vascular images close to surface, OCTA and LASCI are not ideal for detecting hemodynamics hidden underneath densely scattering tissue. A primary all-optical technique to non-invasively detect dynamic events deep inside tissue is diffuse correlation spectroscopy (DCS)~\cite{durduran2014diffuse}. DCS detects hemodynamic events by recording the decorrelation of the light: when coherent light enters thick turbid media, such as tissue, it randomly scatters and produces a speckle pattern. Living tissue is full of microscopic movements, which causes the light to fluctuate, or decorrelate~\cite{brake2016analyzing}. Different phenomena (e.g., tissue movement or blood flow) occur at different speeds, which causes the rate of light decorrelation to differ. In the past, DCS has been widely applied to study brain activity and cerebral health by monitoring cerebral blood flow~\cite{buckley2014diffuse}. To probe deep inside tissue, DCS needs to sample the fluctuations of a few speckle modes at a very high speed (microsecond sampling periods). Thus, traditional implementations usually use only one or very few fibers to collect light from the surface, with the light from each fiber detected by one or few single-pixel single photon sensitive detectors, such as single photon avalanche detectors (SPADs), or photomultipler tubes (PMTs). However, detecting light from only one surface location limits localization accuracy. Moreover, few photons per speckle mode reach the surface after traveling through highly turbid media. To achieve a sufficient signal-to-noise ratio (SNR), long integration times are thus required to achieve a useful estimation of the light decorrelation, which limits the ability to detect transient biological events. While the previous methods can mechanically translate the DCS probe to measure speckles from different surface locations to improve spatial localization~\cite{han2015non,he2015noncontact}, this further increases the data acquisition time, the risk of motion-induced artifacts, and setup complexity.

\begin{figure}[!t]
\begin{center}
\includegraphics[width=15cm]{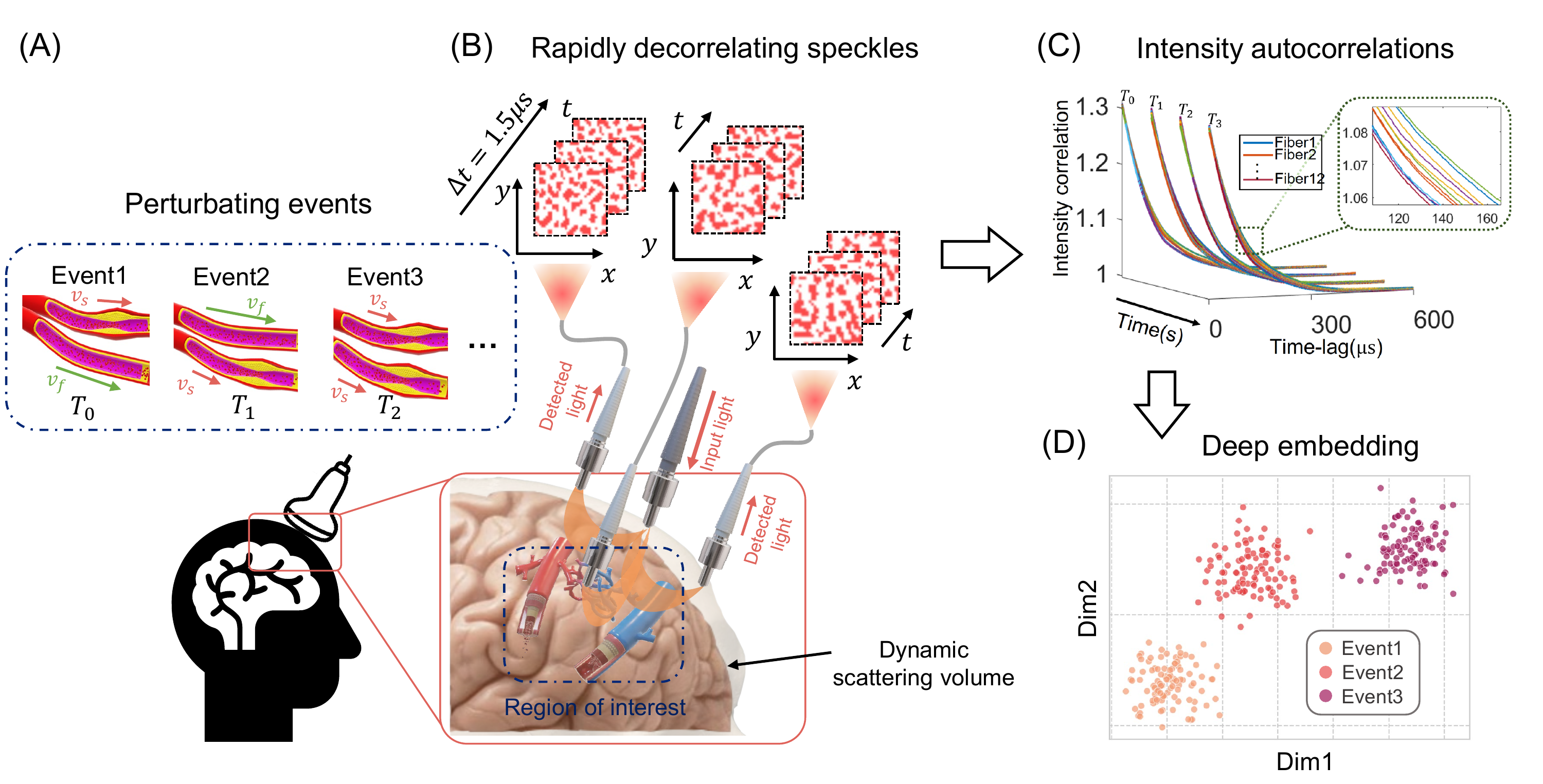}
\end{center}
\caption{Overview of the proposed CREPE technique for classification of events occurring many millimeters within tissue. (A) Different decorrelation phenomena (e.g., different blood vessels flowing at different speeds) deep inside tissue cause surface light speckles in (B) to change at different rates. Speckle fluctuations are collected by fibers and recorded by a SPAD array camera. Temporal intensity autocorrelations of each fiber position for every decorrelation event are computed (C) and classified (D) into different categories using a deep clustering network.}
\label{fig::fig1_overview}
\end{figure}

Recently developed highly parallelized DCS (PaDS) demonstrates that detecting multiple speckles across many optical sensor pixels results in significantly faster correlation sampling rate~\cite{johansson2019multipixel,liu2020fast,liu2020classifying,liu2021fast,sie2020high,xu2021imaging,zhou2021functional,xu2021rapid,xu2022speckle}. Further, advances in contrastive representation learning ~\cite{liu2021self} facilitates the use of deep artificial neural networks to create an embedding space where similar inputs of unique sub-types are clustered together without any data labeling required. As training ground truth labels are usually expensive to acquire in experiments, it is strongly desired to adopt a deep contrastive learning method that works well with unsupervised data ~\cite{yang2017towards}. Building upon these insights, we propose a new technique here, termed \emph{\textbf{C}lassifying \textbf{R}apid decorrelation \textbf{E}vents via \textbf{P}arallelized single photon d\textbf{E}tection (CREPE)}, which uses a novel multi-fiber PaDS system based on massive parallel detection using a 32×32 SPAD array. Figure \ref{fig::fig1_overview} provides a conceptual illustration of the proposed method. The key features are
\begin{itemize}
\item The highly parallelized light detection improves the SNR and sensitivity of the DCS, and detecting speckles from multiple surface positions allows localizing and classifying spatiotemporally varying decorrelating patterns. 
\item CREPE is a \emph{zero-shot} method, meaning it does \emph{not} require training with labels or external datasets \cite{xian2017zero,hospedales2020meta}. 
\end{itemize}
We validate this novel methodology by accurately classifying spatiotemporally varying patterns hidden beneath a 5mm tissue-like phantom made with rapidly decorrelating scattering media.

\section{Method}
\subsection{Tissue phantom design}
\label{sec::tissue}
\begin{figure}[!t]
\begin{center}
\includegraphics[width=15cm]{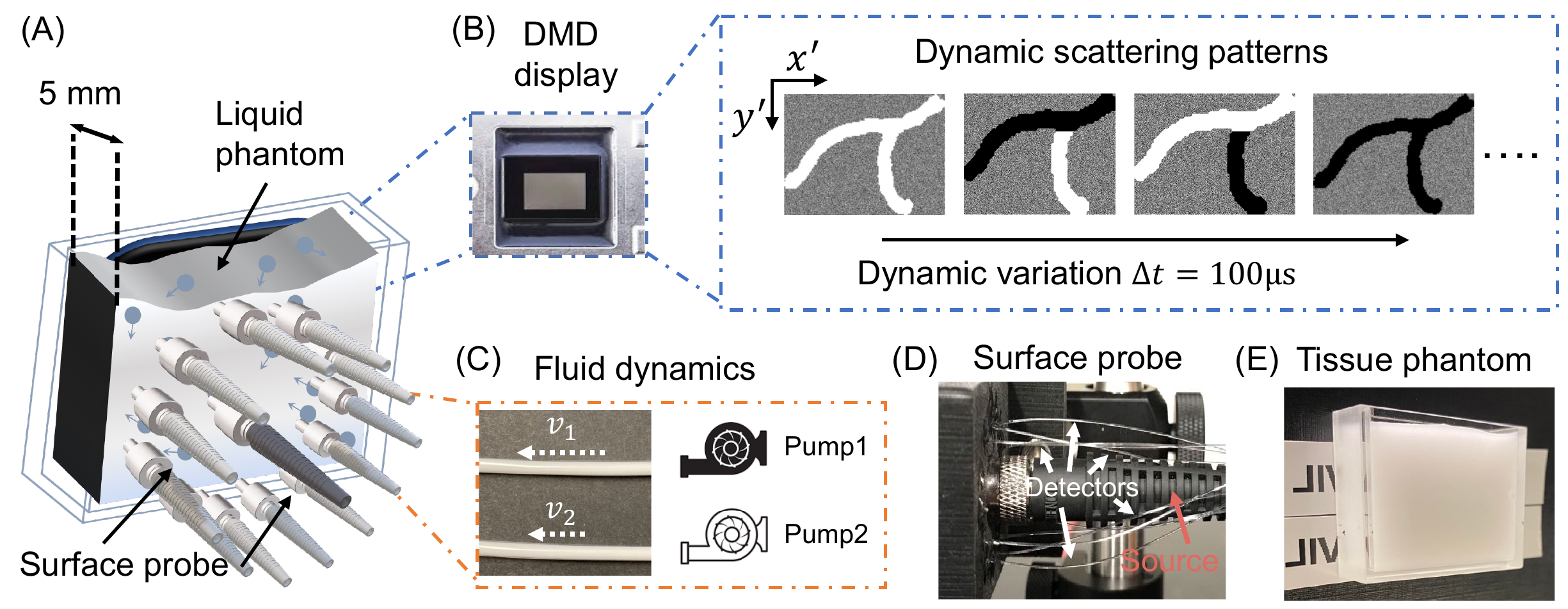}
\end{center}
\caption{(A) illustrates the tissue phantom used. Light from 12 different surface positions were collected with multimode fibers placed circularly around the center source. (B)-(C) are two different mechanisms we used to generate perturbed decorrelation phenomena. (B) A DMD pane hidden underneath the liquid phantom, flipping at multi-kilo hertz rate. (C) Scattering liquid contained in 3mm transparent plastic tubes, flowing at constant speeds. (D-E) are photos of PaDS probe and the tissue phantom used.}
\label{fig::fig2_sche}
\end{figure}

Figure \ref{fig::fig2_sche}(A-C) illustrates our phantom setup. To create dynamic scattering phantoms that mimic movements within living tissue, we used polysterene microsphere solutions at two different concentrations($4.55\times10^6$\#/mm$^3$ and $7.58\times10^6$\#/mm$^3$) enclosed in a thin-walled 5-mm thick cuvette. We termed these two scattering volumes as Tissue I and Tissue II, which results in an estimated reduced scattering coefficient of $\mu_s'=0.7\text{mm}^{-1}$ and experimentally measured absorption coefficient of $\mu_a=0.01\text{mm}^{-1}$ for Tissue I, and $\mu_s'=1.2\text{mm}^{-1},\mu_s'=0.02\text{mm}^{-1}$ for Tissue II~\cite{liu2021fast}. These optical properties closely resemble the optical properties of tissue from  human and model organisms, respectively~\cite{durduran2010diffuse,jacques2013optical}. Underneath the tissue phantom, we placed dynamically fluctuating objects that perturb the decorrelation measured at the surface. We considered two different decorrelation perturbation mechanisms. First, we used a fast changing DMD display flipping at multi-kilo-hertz. We used such display as it's easily reconfigurable and can generate various spatial-temporal varying dynamic scattering patterns that induce additional decorrelation similar to biological phenomena, such as blood flow~\cite{liu2021fast}. Second, we placed two plastic tubes containing the same solution flowing at constant rates. The speed of the flowing liquid inside the tube was controlled with two syringe drivers (New Era, US1010). While this is not as versatile as the DMD, in this way we were able to create more biologically realistic events by mimicking blood vessels. To measure the light fluctuation from different surface locations, we used a 12-fiber-detector PaDS system carefully described in ~\cite{xu2021diffusing}. Figure \ref{fig::fig2_sche}(D-E) presents picture of the PaDS probe and tissue phantom we used. 
\subsection{Data processing}
\begin{figure}[!t]
\begin{center}
\includegraphics[width=15cm]{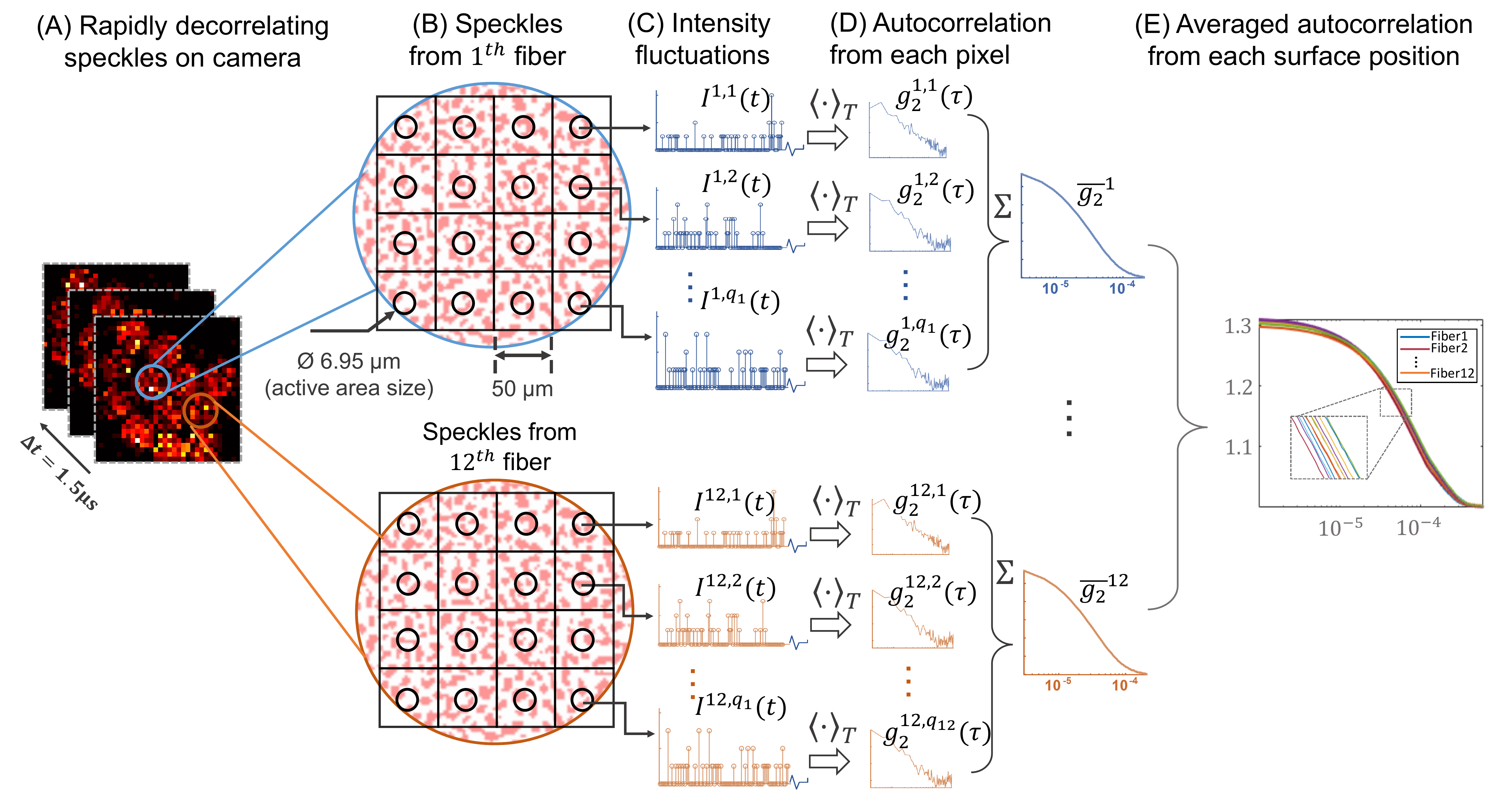}
\end{center}
\caption{Method for computing the autocorrelation curves detailed in \textit{Data processing} subsection. (A) shows a few representative frames captured with the SPAD array. (B)-(E) shows the data processing method, where the autocorrelation from each SPAD pixel were computed, and averaged across each fiber position to generate a set of curves for each decorrelating event.}
\label{fig::fig3_decor}
\end{figure}

To generate a data point per decorrelation event, the temporal autocorrelation for each fiber location was estimated. Although there are other ways to compute temporal statistics across a SPAD array~\cite{valdes2014speckle,jazani2019alternative}), this per-pixel method is robust and widely used~\cite{johansson2019multipixel,sie2020high,liu2021fast}. Figure \ref{fig::fig3_decor}(a) illustrates several representative frames captured by the SPAD camera, sampling at 667kHz (1.5us sampling period), in which the speckles in each pixel fluctuated rapidly. We first computed the normalized temporal intensity autocorrelation ~\cite{durduran2014diffuse} of each pixel as
\begin{linenomath}
\begin{equation}
    g_2^{p,q}(\tau) = \frac{\langle I^{p,q}(t)I^{p,q}(t+\tau)\rangle_{T_{int}}}{\langle I^{p,q}(t){\rangle}_{T_{int}}^2},
\label{eq:correlation}
\end{equation}
\end{linenomath}

\begin{figure}[!t]
\begin{center}
\includegraphics[width=15cm]{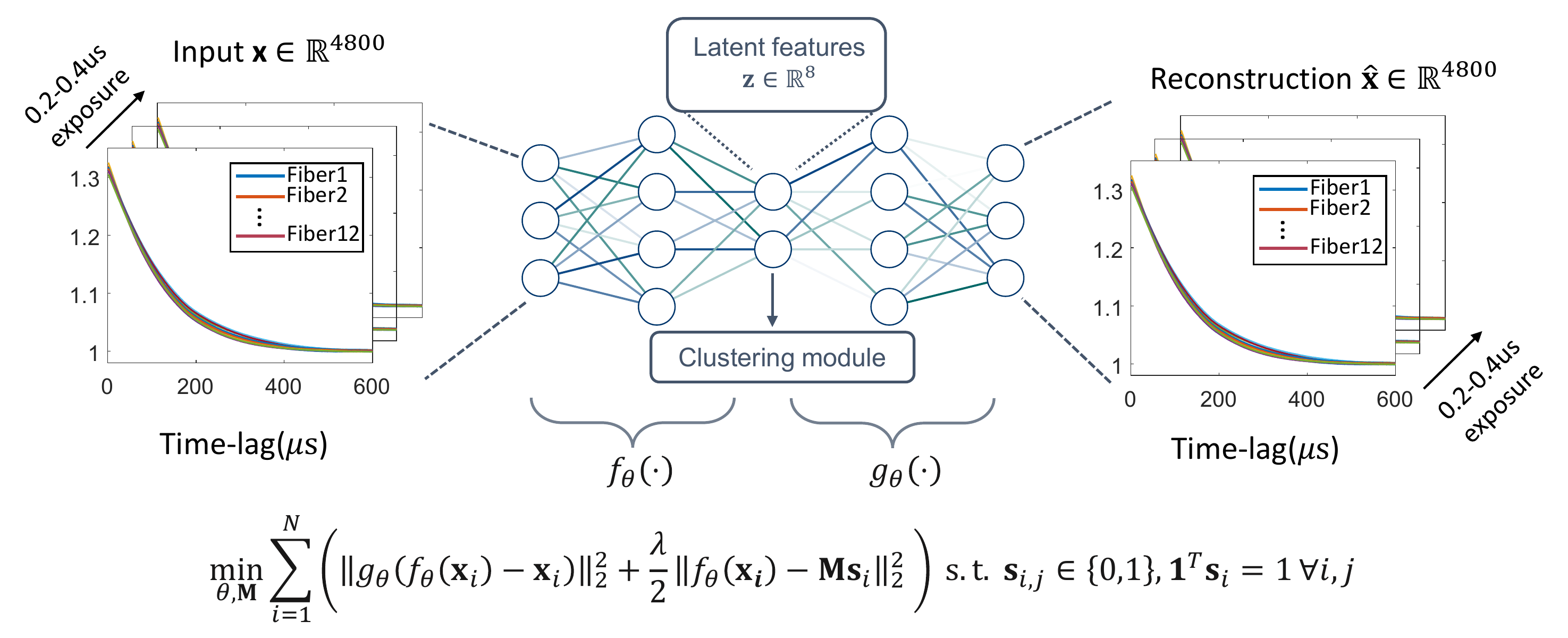}
\end{center}
\caption{Proposed deep clustering method for zero-shot decorrelation event classification. The network contains a stacked auto-encoder that transfers the input data into a latent low-dimension space, then reconstructs the input data from the latent features. A clustering module is used to impact the network weights update to form a classification friendly low-dimension space. Overall, the network is trained with the loss function at the bottom of the figure, with all the variables explained at the end of \textit{Data processing} subsection.}
\label{fig::fig4_nn}
\end{figure}
where $I^{p,q}(t)$ is the number of photons detected by the $q$-th SPAD for $p$-th fiber at time $t$; $\tau$ is the time delay, and ${\langle}\cdot{\rangle}_{T_{int}}$ computes time-average estimated by integrating over $T_{int}$. After calculating $g_2^{p,q}(\tau)$ for every single SPAD, we can obtain an ensemble-averaged, noise-reduced autocorrelation $\overline{g_2^{p}(\tau)}$ for each fiber position by averaging $g_2^{p,q}(\tau)$ that are collected by the $Q_p$ unique SPADs detecting light emitted by the same multi-mode detection fiber, 
\begin{linenomath}
\begin{equation}
    \overline{g}_2^{p}(\tau)=\frac{1}{Q_p}\sum_{q=1}^{Q_p}g_2^{p,q}
\label{eq:average_corr}
\end{equation}
\end{linenomath}
for the $p^{th}$ multi-mode fiber (MMF). We use a look-up table to identify the $Q_p$ SPADs within the array that receives light from the $p$th MMF. Next, we compile the $g_2^{p}(\tau)$ from each fiber into a set of 12 average intensity autocorrelation curves per decorrelation event,  $\{\vx_i\}_{i=1,2,..,N}$, for $N$ events of interest, and aim to classify these event measurements into $K$ categories. While one could use a simple clustering method such as k-means, the high dimensionality inherent to PaDS data benefits from dimensionality reduction. Recent advances in deep unsupervised learning demonstrate that a non-linear transform, such as an artificial neural network, can generate clustering-friendly embedding for state-of-the-art classification results when jointly trained with the cluster module ~\cite{aljalbout2018clustering}. Therefore, we proposed to use a deep clustering network (DCN)~\cite{yang2017towards} to learn a low-dimension representation of the PaDS data for classification, as detailed in Fig.\ref{fig::fig4_nn}. The DCN contains a stacked autoencoder, consisting of an encoder $f_\theta(\cdot)$ that embeds the PaDS data into a low-dimension manifold before a decoder $g_\theta(\cdot)$ maps the embedding back to the original space of the data point. A k-means++ clustering module ~\cite{arthur2006k} is connected to the dimension-reduced latent features of the network, aiming to help weights update to separate the data points in the low-dimension space. Mathematically, the problem can be formulated by the cost function
\begin{linenomath}
\begin{equation}
    \min\limits_{\theta,\mM}\sum_{i=1}^{N} \Big(\|g_{\mathbf{\theta}}(f_{\mathbf{\theta}}(\vx_i))-\vx_i\|_2^2+\frac{\lambda}{2}\|f_{\mathbf{\theta}}(\vx_i)-\mM\mathbf{s}_i\|_2^2\Big) \;\; \text{s.t.} \;\; \mathbf{s}_{i,j} \in \{0,1\}, \mathbf{1}^T\mathbf{s}_i = 1 \forall i,j, 
\end{equation}
\end{linenomath}
where $\vs_i$ is the one-hot assignment vector for $\vx_i$, picking up one-column from $\mM$.  The $k$-th column of $\mM$ represents the centroid of the $k$-th cluster. $\vs_{i,j}$ stands for the $j$-th element of $\vs_i$. The first $\ell_2$ loss here is the data fidelity term, which ensures the ``bottleneck'' contains information to reconstruct the high-dimension autocorrelation curves. The contrastive k-means clustering-specific loss help separate the data points in the embedding space. To jointly optimize the two parts of loss, we alternate between updating the autoencoder weights using stochastic gradient, and finding new centroids for clusters. 

\section{Results}
\begin{figure}[!t]
\begin{center}
\includegraphics[width=14cm]{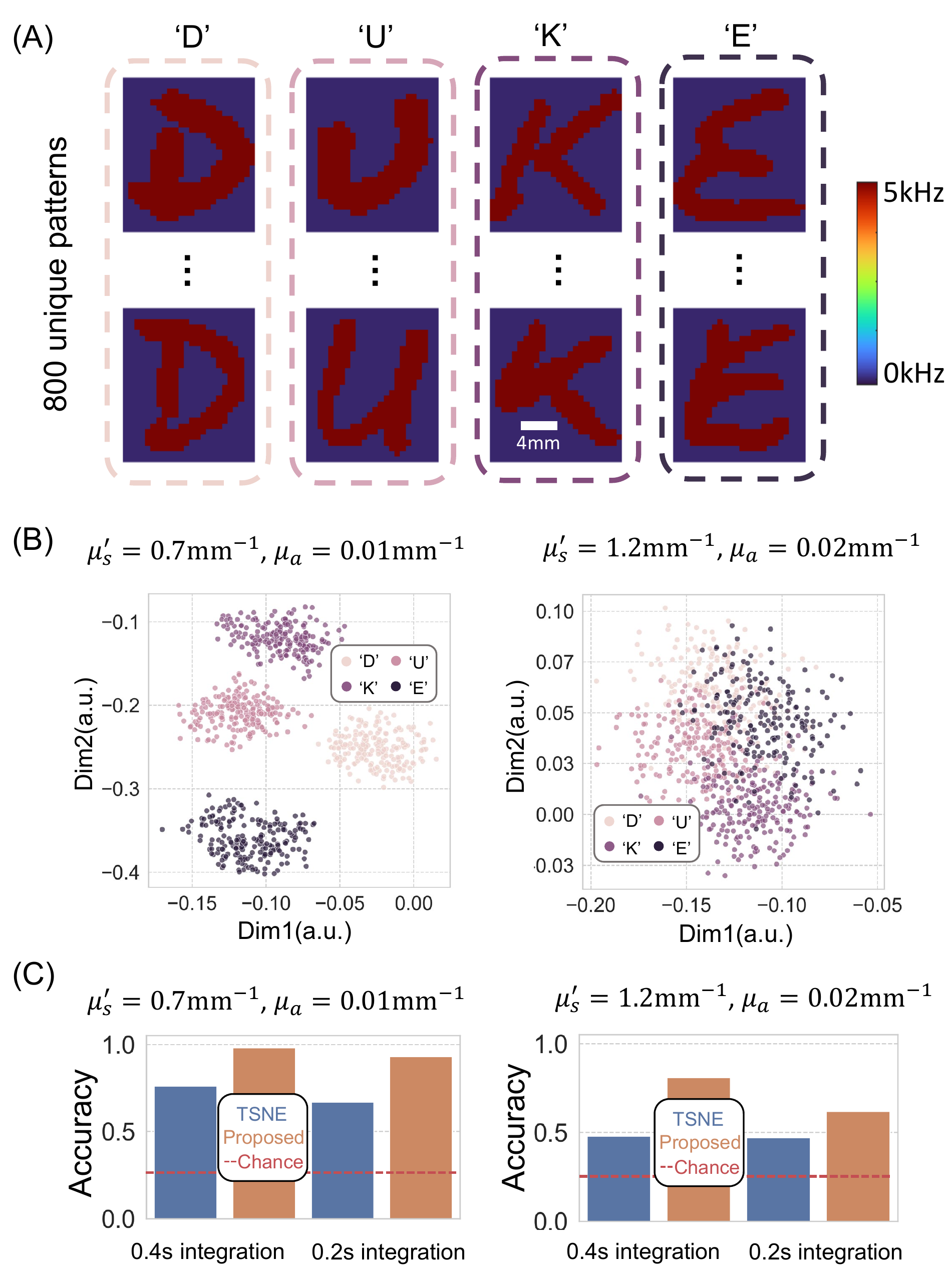}
\end{center}
\caption{(A) depicts some representative spatially different letter-shaped decorrelation events we attempt to classify. The perturbed decorrelations are generated by flipping the DMD at 5 kilo hertz. (B) plots two of the eight dimensions of the embedding using our proposed method. The autocorrelations were computed using 0.4s integration time. (C) barplots of the classification accuracy of TSNE and proposed method using 0.2s and 0.4s integration time. The red dashed line plots the baseline by random guess, which is 0.25 for a quaternary classification task.}
\label{fig::fig5_duke}
\end{figure}
\begin{figure}[!t]
\begin{center}
\includegraphics[width=14cm]{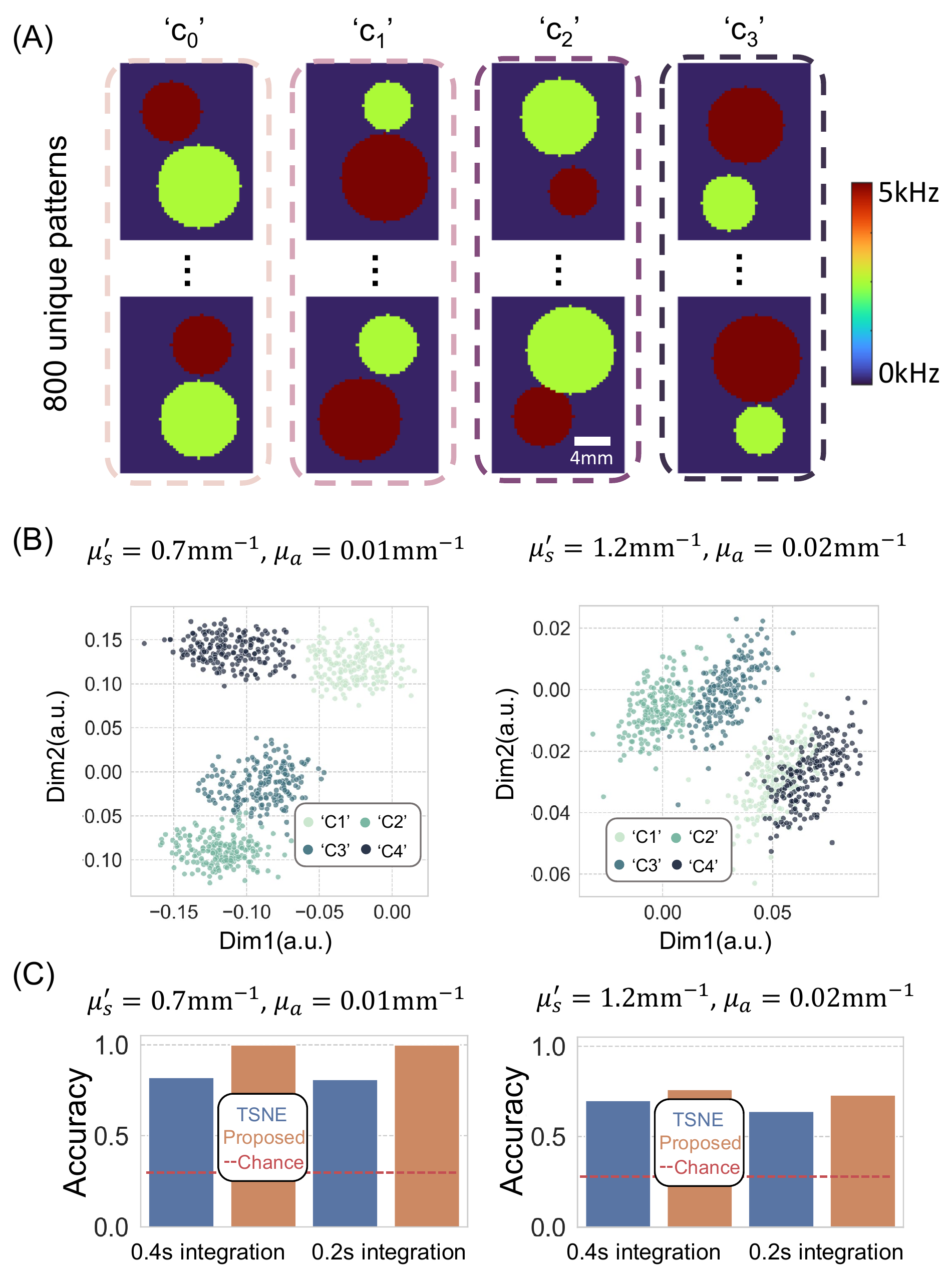}
\end{center}
\caption{(A) depicts some representative spatio-temporally differing circular-shaped decorrelation events we attempted to classify. The perturbed decorrelations were generated by flipping the DMD at 5-10 kilo hertz. (B) plots two of the eight dimensions of the embedding using proposed method. The autocorrelations were computed using 0.4s integration time. (C) Barplots of the classification accuracy of TSNE and proposed method using 0.2s and 0.4s integration times. The red dashed line plots the baseline of chance, which is 0.25 for a quaternary classification task.}
\label{fig::fig6_twocirc}
\end{figure}
\begin{figure}[!t]
\begin{center}
\includegraphics[width=15cm]{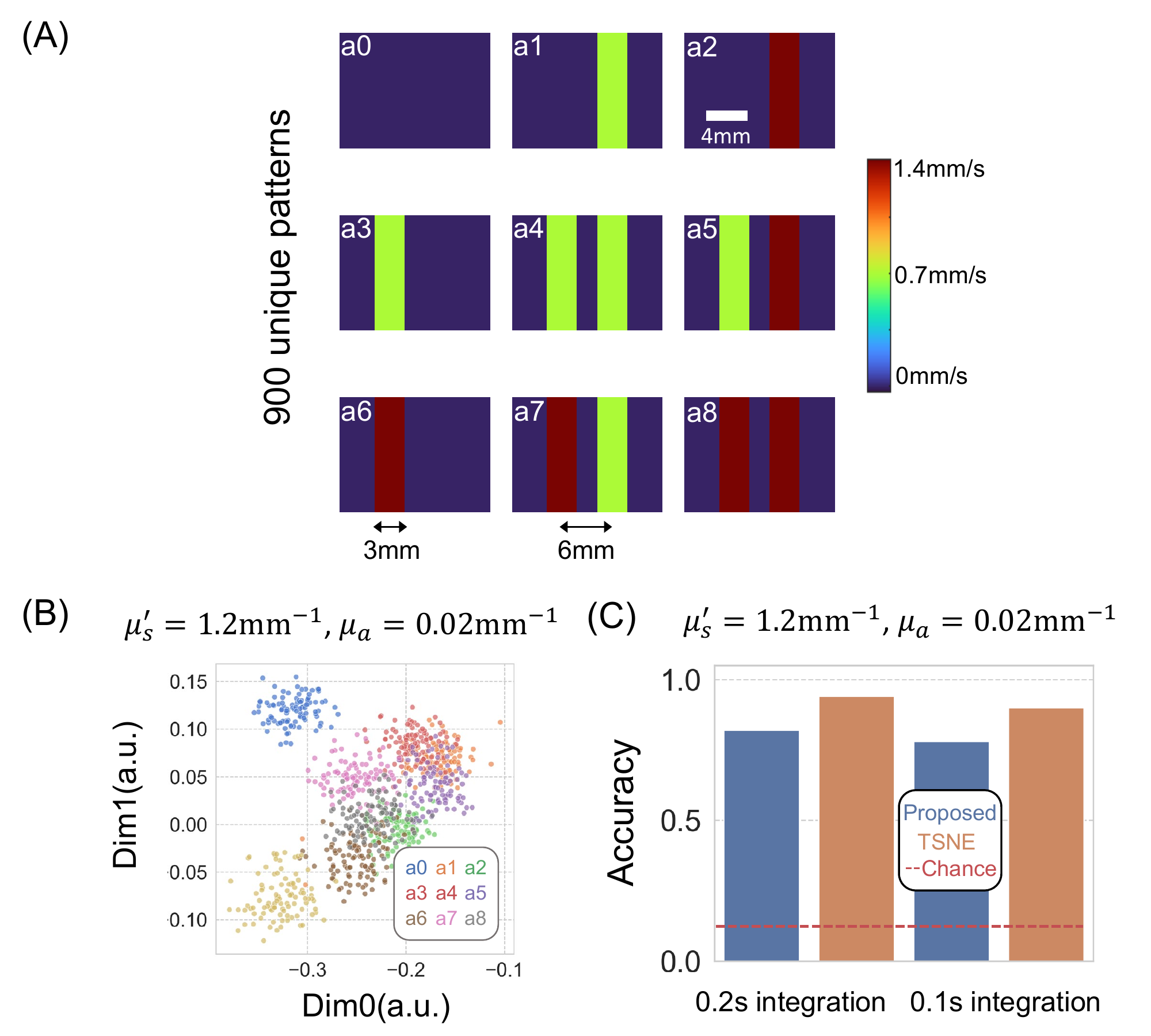}
\end{center}
\caption{(A) depicts the nine decorrelation events we attempted to separate. Those events were generated by placing two 3mm diameter tubes filled with scattering volume placed underneath the liquid phantom. The scattering liquid in the tube either did not flow, or flowed at 1.4mm/s and 0.7mm/s, driven by two syringe pumps. (B) plots two of the eight dimensions of the embedding using TSNE and proposed method. (C) Barplots of the classification accuracy of TSNE and proposed method using 0.1s and 0.2s integration time. The red dashed line plots the baseline of chance, which is 0.11 for a nine-category classification task.}
\label{fig::fig7_syringe}
\end{figure}
We created three datasets as a first validation of our new method, to evaluate the performance in separating spatial, temporal, and spatio-temporal varying decorrelating events. We first displayed 800 spatially different patterns, in this case, handwritten letters from the EMNIST dataset (4 classes: ``D'', ``U'', ``K'', ``E''; 200 examples of each) onto the $10.6\times13.9\text{mm}^2$ fixed DMD area. Some representative patterns are shown in Fig.\ref{fig::fig5_duke}(A). We attempted to separate these decorrelation patterns into their categories using both proposed DCN method and t-distributed stochastic neighbor embedding (TSNE)~\cite{van2008visualizing}, a widely used classic dimension reduction method. The decorrelation patterns were placed underneath 5mm turbid volume described in \emph{Tissue phantom design} subsection. Figure \ref{fig::fig5_duke} (B) plots two of the eight reduced-dimensions from the 800 events using proposed method. These data points were generated by decorrelation events hidden under 5mm turbid volume and the autocorrelations were computed using a 0.4s integration time. Figure \ref{fig::fig5_duke} (C) summarizes the classification accuracy of both methods at two different integration times. We see that both methods (TSNE and proposed) can classify the decorrelation events with accuracy higher than chance (25\% accuracy for quaternary classifications), but the proposed method performs better. We note that the classification accuracy for events hidden beneath Tissue I ($\mu_s'=1.2\text{mm}^{-1}, \mu_a=0.02\text{mm}^{-1}$, close to human tissue optical property) are lower than for Tissue II ($\mu_s'=0.7\text{mm}^{-1}, \mu_a=0.01\text{mm}^{-1}$, close to model organisms tissue properties). This is because the sensitivity of our PaDS method in detecting fast, small decorrelation events decreases as the scattering scene becomes more turbid~\cite{liu2021fast}. Additionally, while reduced integration allows identification of more transient events, the accuracy when using 0.2s integration time is less than when using 0.4s.

Next, we presented 800 spatio-temporally varying patterns containing two differently sized circles onto the DMD display (as shown in Fig.\ref{fig::fig6_twocirc}(A)). Similarly, we plotted two of the eight reduced dimensions using both TSNE and proposed method. Again, these data points were generated by computing the autocorrelations using 0.4s integration time. We see the method performs better at classifying two circles of different sizes and speeds than classifying the letters, due to the fact that the perturbed decorrelation areas covered by the two circles are larger than the those of the letters.

Finally, we applied our method to classify temporally varying patterns generated using two 3mm tubes (Fig.\ref{fig::fig7_syringe}(A)). The dynamic scattering fluid in the tubes either did not flow, or flowed at 1.4mm/sec and 0.7mm/sec (as reference, human arterial blood flow at 4.9-19 cm/sec, while venous blood flow at 1.5-7.1 cm/sec ~\cite{klarhofer2001high}), driven by two syringe pumps. This resulted in nine different possible combinations (Fig.\ref{fig::fig7_syringe}(A)). We generated 100 decorrelation events for each category, resulting in 900 data points. As the perturbations generated using fluid dynamics were more noticeable than the DMD, we only show results using Tissue II. Figure \ref{fig::fig7_syringe} (B) plots two of the eight reduced dimensions of the 900 data points using both methods at 0.2s integration time. Figure \ref{fig::fig7_syringe} (C) summarizes the accuracy of both methods using 0.1 and 0.2s integration time. 

\section{Discussion}
In summary, we developed CREPE, a parallelized, fast, sensitive photon sensing method that records the speckle fluctuations from 12 unique tissue surface positions, along with a deep embedding processing software that can separate the decorrelation events occurring underneath turbid volumes. As a first demonstration, we showed that our approach can detect and categorize various transient movement perturbations through rapidly decorrelating dynamic scattering tissue phantoms. Our method does not require expensive data labels to train the network, and therefore has a great potential to be applied in clinical \emph{in vivo} studies. To ensure effective clinical translation, there are several improvements that can be made to both the system design and processing algorithm. First, as shown in camera images in Fig.\ref{fig::fig3_decor}, the detection fiber bundle we use did not map surface speckles to all $32\times32$ SPAD pixels to maximize the speckle detection efficiency. Future work should strive to custom-design a fiber bundle that provides better array coverage. In addition, to cover deeper regions of tissue, longer source-detector separation is desired. While it is difficult to further increase the SPAD array sampling rate, which is required to record light traveling longer distances, 
we expect pixel-count for monolithic CMOS SPAD arrays to continue to rise (e.g., one megapixel SPAD arrays are now available~\cite{canon2021mega}). This provides promising opportunities to utilize spatial speckle statistics to help understand decorrelation events occurring deep in tissue~\cite{valdes2014speckle,xu2022speckle}. Integrating CREPE with these speckle contrast methods on a SPAD array with higher pixel counts should be investigated to ensure reliable translation into clinical use.

\section*{Author Contributions}
S. X., W. L., X. Y., R. Q., K. K. and P. C. K. constructed the hardware setup. S. X., W. L. and J. J. designed the software. S. X, K. Z, L. K, E. B., P. M. and R. H wrote the manuscript. H. W., E. B., S. H. and R. H. supervised the project. 

\section*{Funding}
Research reported in this publication was supported by the National Institute of Neurological Disorders and Stroke of the National Institutes of Health under award number RF1NS113287, as well as the Duke-Coulter Translational Partnership. W.L. acknowledges the support from the China Scholarship Council. R.H. acknowledges support from a Hartwell Foundation Individual Biomedical Researcher Award, and Air Force Office of Scientific Research under award number FA9550. 

\section*{Acknowledgments}
The authors also want to thank Kernel Inc. for their generous support. In addition, the authors would like to express our great appreciation to Dr. Haowen Ruan for inspirational discussion.

\section*{Data Availability Statement}
The data support the findings of this study are available from the corresponding author through collaborative investigations and upon reasonable request.
\section*{Conflict of interest} S.X. and R.H. have submitted a patent application related to this work, assigned to Duke University
\bibliographystyle{unsrt}  
\bibliography{test}


\end{document}